\documentclass[aps,showkeys,showpacs,twocolumn,prd]{revtex4}
\usepackage{graphicx}
\usepackage{amsfonts}
\usepackage{graphicx}
\usepackage{graphics}
\usepackage[anne,small]{caption}
\usepackage{subfig}
\usepackage{bm}
\usepackage{dcolumn}
\usepackage{url}

\def\bea{\begin{eqnarray}}
\def\eea{\end{eqnarray}}
\def\bJ{\bm{J}}
\def\bx{\bm{x}}
\def\bphi{\bm{\phi}}
\def\vecr{\bm{\hat{r}}}
\def\bomega{\bm{\omega}}
\def\rmd{\text{d}}

\begin{document}
\title{Spontaneous Breaking of Rotational Symmetry in Rotating Solitons -- \\
a Toy Model of Excited Nucleons with High Angular Momentum}
\date{\today}
\author{Itay Hen}
\email{itayhe@post.tau.ac.il}
\author{Marek Karliner}
\email{marek@proton.tau.ac.il}
\affiliation{Raymond and Beverly Sackler School of Physics and Astronomy 
Tel-Aviv University, Tel-Aviv 69978, Israel}
\begin{abstract}
We study the phenomenon of spontaneous breaking of rotational symmetry (SBRS) in the rotating
solutions of two types of baby Skyrme models. In the first the domain is a two-sphere
and in the other, the Skyrmions are confined to the interior of a unit disk. 
Numerical full-field results show that when the angular momentum
of the Skyrmions increases above a certain critical value,
the rotational symmetry of the solutions is broken and the minimal energy configurations
become less symmetric. 
We propose a possible mechanism as to why SBRS is present in 
the rotating solutions of these models, while it is not observed in
the `usual' baby Skyrme model.  
Our results might be relevant for a qualitative understanding of 
the non-spherical deformation of 
excited nucleons with high orbital angular momentum.
\end{abstract}

\maketitle

\section{Introduction}
The phenomenon of spontaneous breaking of rotational symmetry (SBRS) in rotating systems
relates to occurrences in which physical systems which rotate fast enough
deform in a manner which breaks their rotational symmetry, 
a symmetry which is present when these systems are static 
or rotating slowly. 
The recognition that rotating physical systems 
can yield solutions with less symmetry than the governing equations 
is not new. One famous example which dates back to $1834$ is that of
the equilibrium configurations of a rotating fluid mass.  It was  
Jacobi who was first to discover that,
if rotated fast enough, a self-gravitating fluid mass can have equilibrium
configurations lacking rotational symmetry.
In modern terminology, Jacobi's asymmetric equilibria appear through a symmetry breaking 
bifurcation from a family of symmetric equilibria as the angular momentum 
of the system increases above a critical value (a ``bifurcation point'') \cite{Lyttle,Chandra}.
Above this critical value, rotationally-symmetric configurations 
are no longer stable, and configurations with a broken rotational symmetry become
energetically favorable.
\par
By now it is widely recognized that symmetry-breaking bifurcations 
in rotating systems are of frequent occurrence and that this is in fact a very general phenomenon,
appearing in a variety of physical settings among which are
fluid dynamics, star formation, heavy nuclei, chemical reactions, plasmas, 
and biological systems, to mention some diverse examples. 
\par
Recently, SBRS has also been observed in self-gravitating
$N$-body systems \cite{Votyakov1,Votyakov2}, where 
the equilibrium configurations of an $N$-body self-gravitating system enclosed in a finite 3 dimensional spherical volume
have been investigated using a mean-field approach. 
It was shown that when the ratio of the angular momentum of the system to its energy is high, 
spontaneous breaking of rotational symmetry occurs, manifesting itself in the formation of double-cluster
structures. These results have also been confirmed with direct numerical simulations \cite{LMS}.
\par
It is well-known that a large number of phenomena exhibited by many-body systems have
their counterparts and parallels in field theory, which in some sense is
a limiting case of $N$-body systems in the limit $N \to \infty$. 
Since the closest analogies to a lump of matter in field theories are solitons,
the presence of SBRS in self-gravitating $N$-body systems has led us to expect that it 
may also be present in solitonic field theories.
\par
Our main motivation towards studying SBRS in solitons
is that in hadronic physics Skyrme-type solitons \cite{Zahed:1986qz}
often provide a fairly
good qualitative description of nucleon properties. 
In particular, it is interesting to ask what
happens when such solitons rotate quickly,
because this might shed some light on the non-spherical deformation of 
excited nucleons with high orbital angular momentum, a subject which is now
of considerable interest. We address this issue in more detail in the concluding section 
of this manuscript. 

In what follows, we study SBRS in one
of the simplest and well-known field theoretic models
admitting stable rotating solitonic solutions,
namely the baby Skyrme model \cite{Old1,Old2}.
First, we give a brief account for 
the occurrence of SBRS in physical systems in general,
and then use the insights 
gained from this discussion to infer the conditions under which SBRS might appear 
in solitonic models and in that context study its appearance in baby Skyrme models.
Specifically, we shall show that SBRS emerges if the domain manifold of the model is
a two-sphere or a disk, while if the domain is $\mathbb{R}^2$,
SBRS does not occur. 

\section{SBRS from a dynamical point of view}
The onset of SBRS may be qualitatively understood as resulting from 
a competition between the static energy of a system and its moment of inertia.
To see this, 
let us consider a system described by a set of degrees of freedom $\phi$,
and assume that the dynamics of the system is governed by a Lagrangian which is invariant under spatial rotations.
When the system is static, its equilibrium configuration
is obtained by minimizing its static energy $E_{\textrm{static}}$ with respect to its degrees of freedom $\phi$
\begin{equation} \label{eq:Estat1}
\frac{\delta E}{\delta \phi}=0 \quad \textrm{where} \quad E=E_{\textrm{static}}(\phi) \,.
\end{equation}
Usually, if $E_{\textrm{static}}(\phi)$ does not include terms which manifestly break rotational
symmetry, the solution to (\ref{eq:Estat1}) is rotationally symmetric
(with the exception of degenerate spontaneously-broken vacua, which are not of our concern 
here).
If the system rotates with a given angular momentum $\bJ=J \hat{z}$, its configuration is
naturally deformed. Assuming that
the Lagrangian of the system is quadratic in the time derivatives,
stable rotating configurations (if such exist) are obtained by minimizing its total
energy  $E_{J}$
\begin{equation} \label{eq:Erot1}
\frac{\delta E_{J}}{\delta \phi}=0 \quad \textrm{where} \quad E_{J}=
E_{\textrm{static}}(\phi)+\frac{J^2}{2 I(\phi)} \,, 
\end{equation}
where $I(\phi)$ is the ratio between the angular momentum of the system and its angular velocity  $\bomega=\omega \hat{z}$
(which for simplicity we assume is oriented in the direction of the angular momentum). 
$I(\phi)$ is the (scalar) moment of inertia of the system.
\par
The energy functional (\ref{eq:Erot1})  consists of two terms.
The first, $E_{\textrm{static}}$,
increases with the asymmetry. This is simply a manifestation of the 
minimal-energy configuration in the static case
being rotationally symmetric. The second term
$J^2/2I$, having the moment of inertia in the denominator,
decreases with the asymmetry. 
At low values of angular momentum, the $E_{\textrm{static}}$ term 
dominates, and thus asymmetry is not energetically favorable, but as the value of angular momentum
increases, it is the second term which becomes dominant, 
thus giving rise to a possible breaking of rotational symmetry.

\subsection{The self gravitating ellipsoid}
As an illustration of the above reasoning let us consider the
simple problem of a self-gravitating ellipsoid of liquid
mass $M$ \cite{diffApp}. 
The density of the ellipsoid  $\rho$ is assumed to be constant but its shape is allowed
to deform. The boundary of the ellipsoid may be parametrized by
\bea \label{eq:Ellipsurface}
&&r(\theta,\varphi)= \\
&&\left( \eta^2 \cos^2 \theta + \frac{4 \pi \rho}{3 M}\frac{\sin^2 \theta}{\eta (1-\epsilon^2)}
\big( 1 +\epsilon^2 -2 \epsilon \cos 2 \varphi \big) 
\right)^{-1/2} \,,\nonumber
\eea
with $\theta \in [0,\pi]$ being the polar angle, and $\varphi  \in [0,2\pi)$ the azimuthal angle.
Here, the ellipsoid has two
degrees of freedom $\phi=(\eta,\epsilon)$, with a third degree of freedom 
eliminated by the constraint of constant volume,
and a non-zero value of $\epsilon$ indicates breaking of rotational symmetry. 
The static energy of the ellipsoid is due to self-gravitation and is given by 
\bea \label{eq:EllipStatic}
&&E_{\textrm{static}}(\eta,\epsilon) = \\
&&- \frac{3}{10} G M^2 \int_0^{\infty} \left((a_1+u)(a_2+u)(a_3+u) \right)^{-1/2} ud \,,\nonumber
\eea
where $\displaystyle{a_1=\frac{3 M \eta}{4 \pi \rho} \frac{1+\epsilon}{1-\epsilon}}$, 
$\displaystyle{a_2=\frac{3 M \eta}{4 \pi \rho} \frac{1-\epsilon}{1+\epsilon}}$ and $a_3=1/\eta^2$ \cite{Lyttle}.
The minimal-energy configuration of the static self-gravitating ellipsoid
is obtained by minimizing (\ref{eq:EllipStatic}) with respect to the parameters $\eta$ and $\epsilon$,
giving
\begin{equation} \label{eq:SolStatic}
\eta=(\frac{4 \rho}{3 \pi M})^{1/3}, \quad \epsilon=0 \quad \to \quad r(\theta,\varphi)=1/\eta. 
\end{equation}
This means that the configuration that minimizes $E_{\textrm{static}}$ 
is a sphere.
When the ellipsoid is rotated with angular momentum $\bJ=J \hat{z}$, the expression
for its energy becomes
\begin{equation} \label{eq:EllipsErot}
E_{J}=E_{\textrm{static}}(\eta,\epsilon)+\frac{J^2}{2 I(\eta,\epsilon)} \,,
\end{equation}
where $I(\eta,\epsilon)$ is the moment of inertia of the ellipsoid
\begin{equation} \label{eq:EllipsMoi}
I(\eta,\epsilon)=\frac{3 M^2 \eta}{10 \pi \rho}  \frac{1+\epsilon^2}{1-\epsilon^2}  \, .
\end{equation}
Note that both $E_{\textrm{static}}(\eta,\epsilon)$ and $I(\eta,\epsilon)$ are monotonically increasing functions
of the symmetry-breaking parameter $\epsilon$, as discussed earlier.
It is the `competition' between these
two expressions in the minimization of (\ref{eq:EllipsErot}), that determines whether and when SBRS 
occurs. 
\par
\begin{figure}[ht!]
\includegraphics[width=0.5\textwidth]{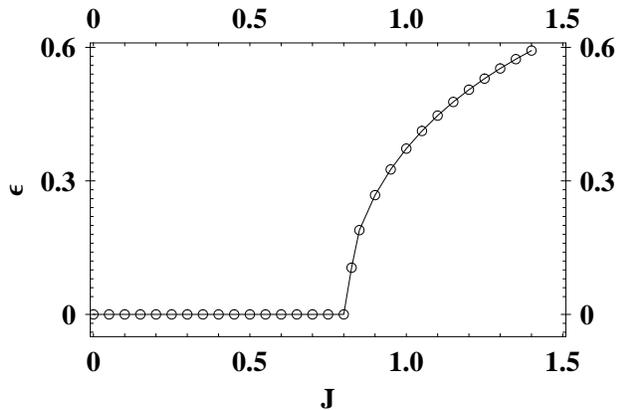}
\caption{\label{gravEllips}The self-gravitating ellipsoid ($M=\frac{4}{3} \pi$, $\rho=1$ and $ G M^2=5/3$):
The `symmetry-breaking' parameter $\epsilon$ for the minimal energy configuration as a function of the angular momentum $J$, showing the existence of 
a critical angular momentum $J_{\textrm{crit}} \approx 0.8$ above which the ellipsoid 
is no longer rotationally symmetric. The line is to guide the eye.$\hfill$}
\end{figure}

A (numerical) minimization 
of the energy functional (\ref{eq:EllipsErot}) for different values of $J$ with respect to
the parameters $\eta$ and $\epsilon$ (the constants
of the problem are taken to be $M=\frac{4}{3} \pi$, $\rho=1$ and $ G M^2=5/3$) indeed reveals
the presence of SBRS. Below a critical value of angular momentum $J_{\textrm{crit}}$ 
(which here is $J_{\textrm{crit}} \approx 0.8$), 
rotationally-symmetric
configurations are energetically more favorable, and $\epsilon=0$ minimizes the energy.
The ellipsoid boundary is an oblate spheroid. 
Above $J_{\textrm{crit}}$ however, the energy functional is
no longer minimized by $\epsilon=0$ and 
bifurcation occurs; the minimal-energy configurations become ellipsoids
with three unequal axes. These results are summarized in Fig. \ref{gravEllips}.

\section{\label{sec:origBaby} SBRS in baby Skyrme models}
In what follows, we show
that the above mechanism of SBRS
is present in solitonic field theories as well,
specifically in certain types of baby Skyrme models.
\par
The baby Skyrme model is a non-linear theory in $(2+1)$ dimensions 
which has several applications in condensed-matter physics \cite{Condensed}.
The target manifold is a three-dimensional vector 
$\bphi$  with the constraint $\bphi \cdot \bphi=1$. 
The Lagrangian density is given by
\bea \label{eq:BabyLag}
\mathcal{L} &= &\frac1{2} \partial_{\mu} \bphi \cdot \partial^{\mu} \bphi 
- \frac{\kappa^2}{2}\big[(\partial_{\mu} \bphi \cdot \partial^{\mu} \bphi)^2 \\\nonumber
&-& (\partial_{\mu} \bphi \cdot \partial_{\nu} \bphi) 
\cdot (\partial^{\mu} \bphi \cdot \partial^{\nu} \bphi)
\big]
-\mu^2(1-\phi_3) \,,
\eea
equipped with a Minkowski metric.
The first term  in the Lagrangian is the usual kinetic term known from $\sigma$ models.
The second term is fourth order
in derivatives and is the analogue of the Skyrme term in the $(3+1)D$ Skyrme model \cite{Sk1,Sk2}.
The last term is a potential term, which
is introduced to ensure the stability of the solutions \cite{pot}.
Henceforth, we shall refer to this model as the `usual' or `original' baby Skyrme model. 
\par
The existence of stable solutions in this model is a consequence of the nontrivial
topology of the mapping $\mathcal{M}$ of the physical space
into the field space  at a given time,
$\quad \mathcal{M}:  \mathbb{R}^2 \mapsto S^2,\quad$ where 
the physical space $\mathbb{R}^2$  is compactified to $S^2$
by requiring the spatial infinity to be equivalent in
each direction.
The topology which stems from this one-point 
compactification allows the classification of maps into equivalence classes,
each of which has a unique conserved quantity called the topological charge.
\par
The static solutions of the baby Skyrme model (\ref{eq:BabyLag}) have  
rotationally-symmetric energy and charge distributions in the charge-one and charge-two sectors \cite{Old1}. 
The charge-one Skyrmion has an energy peak at its center which drops down exponentially.
The energy distribution of the charge-two Skyrmion has a ring-like peak around its center at some 
characteristic distance.
The rotating solutions of the model have also been previously studied \cite{Old2,rotBaby2}.
It has been found that rotation at low angular velocities slightly deforms
the Skyrmion but it remains rotationally-symmetric.
For larger values of angular velocity, the rotationally-symmetric configuration becomes
unstable but in this case the Skyrmion does not undergo symmetry breaking. 
Its stability is restored through a different mechanism, namely that of radiation. 
The Skyrmion radiates out the excessive energy and angular momentum,
and as a result begins slowing down until it reaches equilibrium at some constant angular velocity,
its core remaining rotationally-symmetric. 
Moreover, if the Skyrme fields are 
restricted to a rotationally symmetric (hedgehog) form,
the critical angular velocity above which the Skyrmion radiates can be obtained analytically. 
It is simply the coefficient of the potential term $\omega_{\textrm{crit}}=\mu$
\cite{Old2}.
Numerical full-field simulations we have conducted show that the Skyrmion
actually begins radiating well below $\omega_{\textrm{crit}}$,
as radiation itself may be non-rotationally-symmetric. 
The Skyrmion's core, however, remains rotationally symmetric for every angular velocity.
\par 
The stabilizing effect of the radiation on the solutions of the model
has lead us to believe that models in which radiation 
is somehow inhibited may turn out to be good candidates for the occurrence of SBRS.
In the present paper we study two such baby Skyrme models. In these models
energy and angular momentum are not allowed to escape 
to infinity through radiation,
and for high enough angular momentum the mechanism responsible for SBRS discussed in the previous section 
takes over, revealing solutions with spontaneously broken rotational symmetry.
\par
The first model we discuss is a baby Skyrme model in which the physical space $\mathbb{R}^2$
is replaced by a unit two-sphere, and in the second model Skyrmions are confined to the inside of a unit circle in
$\mathbb{R}^2$. We compute the  minimal-energy configurations 
of the rotating solutions of both models by applying 
a full-field relaxation method with which exact numerical solutions are  obtained. 
For the baby Skyrme model on the two-sphere we also 
take a more analytical approach using rational maps. 
We discuss these models and the minimization method in more detail in the next section.

\section{\label{sec:2sphere} The baby Skyrme model on the two-sphere}
The first baby Skyrme model we investigate which exhibits SBRS is 
the one for which both the domain and target manifolds are unit two-spheres. 
This model may be thought of as Skyrme's original $3$D model 
once the radial coordinate is integrated out \cite{2Sphere3}. 
As in the usual baby Skyrme model, the Lagrangian density is simply
\bea \label{O3lagrangian}
\mathcal{L}&=&\frac1{2} \partial_{\mu} \bphi \cdot \partial^{\mu} \bphi \\\nonumber
&-& \frac{\kappa^2}{2}\big[ 
(\partial_{\mu} \bphi \cdot \partial^{\mu} \bphi)^2
-(\partial_{\mu} \bphi \cdot \partial_{\nu} \bphi)
(\partial^{\mu} \bphi \cdot \partial^{\nu} \bphi)
\big]\,,
\eea
with the metric $\textrm{d} s^2=\textrm{d} t^2 - \textrm{d} \theta^2 -\sin^2 \theta \,\textrm{d} \varphi^2$,
where $\theta$ is the polar angle $\in [0,\pi]$ and $\varphi$
is the azimuthal angle $\in [0,2 \pi)$.
In this model a potential term is not necessary for the stability of the solutions \cite{2Sphere3}
and thus is omitted. 
The Lagrangian of this model
is invariant under rotations in both the domain and the target spaces,
possessing an $O(3)_{\textrm{domain}} \times O(3)_{\textrm{target}}$ symmetry.
\par
As in the original baby Skyrme model, the relevant homotopy group here
is $\pi_2(S^2)=\mathbb{Z}$, implying that each field configuration is 
characterized by an integer topological charge $B$, the topological degree of
the map $\bphi$, which in spherical coordinates is given  by
\bea \label{eq:O3Bdens}
B= \frac1{4 \pi } \int \rmd \, \Omega 
	\frac{ \bphi \cdot (\partial_{\theta} \bphi \times \partial_{\varphi} \bphi )}{
	\sin \theta} \,,
\eea
where $\rmd \Omega= \sin \theta \, \rmd \theta \, \rmd \varphi$. 
Static solutions within each topological sector are obtained by minimizing the energy 
functional
\bea \label{eq:Estat}
E_{\textrm{static}}&=&\frac1{4 \pi B} \int \rmd \Omega \Big(\frac1{2}(\partial_{\theta} \bphi)^2 
+ \frac1{2} \frac{1}{\sin^2 \theta}( \partial_{\varphi} \bphi)^2 \nonumber \\* 
&+& \frac{\kappa^2}{2} \frac{(\partial_{\theta} \bphi \times \partial_{\varphi} \bphi)^2}{\sin^2 \theta}
\Big) \,,
\eea
where the $(4 \pi B)^{-1}$ factor has been inserted for convenience. 
The static solutions of the model were studied in detail in \cite{2Sphere3} up to charge $B=14$.
These which are rotationally-symmetric are 
the charge-one Skyrmion has an analytic ``hedgehog'' solution  with 
spherically-symmetric energy and charge distributions, and 
the charge-two solution has an axially-symmetric ring-like solution.
\par 
In order to find the stable rotating solutions of the model,
we assume for simplicity that 
any stable solution would rotate
around the axis of angular momentum (which is taken to be the $z$ direction) with some angular velocity $\omega$.
The rotating solutions thus take the form \hbox{$\bphi(\theta,\varphi,t)=\bphi(\theta,\varphi- \omega t)$}. 
The energy functional to be minimized is
\bea \label{eq:Eomega}
E=  E_{\textrm{static}}+ \frac{J^2}{2 I} \,,
\eea
where $I$ is the ratio of the angular momentum of the Skyrmion to its angular velocity, or its 
``moment of inertia'', given by
\bea
I = \frac1{4 \pi B}  \int \rmd \Omega \left((\partial_{\varphi} \bphi)^2 
+\kappa^2 (\partial_{\theta} \bphi \times \partial_{\varphi} \bphi)^2
\right) \,.
\eea
\subsection{The numerical procedure}
Since the Euler-Lagrange equations
derived from the energy functional (\ref{eq:Eomega})
are non-linear \textit{PDE}'s, in general the 
minimal energy configurations can only be obtained 
with the aid of numerical techniques. 
In what follows, we obtain the minimal energy configurations 
which correspond to rotating Skyrmion solutions,
using a full-field relaxation method, in which the domain $S^2$ is discretized 
to a spherical grid -- $100$ grid points for $\theta$ 
and $100$ points for $\varphi$. 
The relaxation process begins by initializing the field triplet $\bphi$
to a rotationally-symmetric configuration  
\bea \label{eq:initConfig}
\bphi_{\textrm{initial}}=(\sin \theta \cos B \varphi,\sin \theta \sin B \varphi,\cos \theta) \,,
\eea
where $B$ is the topological charge of the Skyrmion in question.
The energy of the baby Skyrmion is then minimized by repeating the following steps:
a point $(\theta_m,\varphi_n)$ on the grid is chosen at random, along 
with one of the three components of the field $\bphi(\theta_m,\varphi_n)$.
The chosen component is then shifted by a value $\delta_{\phi}$ chosen uniformly 
from the segment $[-\Delta_{\phi},\Delta_{\phi}]$
where $\Delta_{\phi}=0.1$ initially. The field triplet is then normalized
and the change in energy is calculated.
If the energy decreases, the modification of the field is accepted
and otherwise it is discarded.
The procedure is repeated while the value of 
$\Delta_{\phi}$ is gradually decreased throughout
the procedure. This is done until no further decrease in energy is observed.\par
One undesired feature of this minimization scheme
is that it can get stuck at a local minimum.
This problem can be resolved by using the ``simulated annealing'' algorithm \cite{SA1,SA2},
which in fact has been successfully implemented before,
in obtaining the minimal energy configurations of 
static two and three dimensional Skyrmions \cite{SkyrmeSA}. 
The algorithm comprises of repeated applications of a Metropolis algorithm 
with a gradually decreasing temperature,
based on the fact that when a physical system is slowly cooled down,
reaching thermal equilibrium at each temperature, it will end up in its ground state. 
This algorithm, however, is much more expensive in terms of computer time. 
We therefore employ it only on a representative sample of the parameter space,
just as a check on our results, which correspond to a Metropolis algorithm of zero temperature.
\par
\subsection{Results}
In what follows we present the results obtained by the minimization scheme
described in the previous section to the rotating solutions of the model
in the charge-one and charge-two sectors, which as mentioned above are rotationally-symmetric.
For simplicity, we fix the parameter $\kappa$ at $\kappa^2= 0.01$ 
although other $\kappa$ values were tested as well,
yielding qualitatively similar solutions. 

\subsubsection{Rotating charge-one solutions}
\begin{figure}[ht!]
\includegraphics[angle=0,scale=1,width=0.48\textwidth]{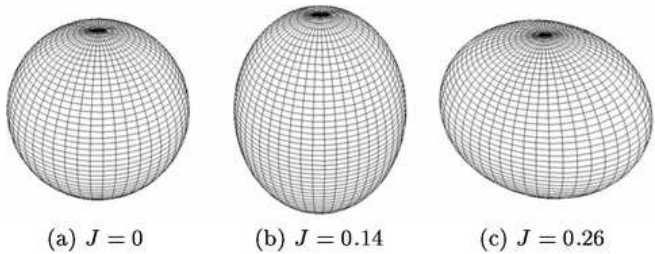}
\caption{\label{O3b1}Baby Skyrmions on the two-sphere ($\kappa^2=0.01$): 
The charge distribution $\mathcal{B}(\theta, \varphi)$ of the charge-one Skyrmion for different angular momenta.
In the figure, the vector 
$\mathcal{B}(\theta, \varphi)\vecr$
is plotted for the various $\theta$ and $\varphi$ values.$\hfill$}
\end{figure}

In perfect analogy with the self-gravitating ellipsoid discussed in
the Introduction, the rotating charge-one Skyrmion, which has spherically-symmetric
energy and charge distributions
in the static limit (Fig. \ref{O3b1}a), was found to exhibit SBRS.
When rotated slowly, its symmetry is reduced to $O(2)$, 
with the axis of symmetry coinciding with the axis of rotation (Fig. \ref{O3b1}b).
At some critical value of angular momentum (which in the current settings is 
$J_{\textrm{crit}} \approx 0.2$), the axial symmetry 
is further broken, yielding an ellipsoidal
energy distribution with three unequal axes (Fig. \ref{O3b1}c). Any further increase in angular momentum
results in the elongation of the Skyrmion in one horizontal direction and its shortening 
in the perpendicular one.  The results are very similar to those
of the rotating self-gravitating ellipsoid.

\subsubsection{Rotating charge-two solutions}
\begin{figure}[ht!]
\includegraphics[angle=0,scale=1,width=0.48\textwidth]{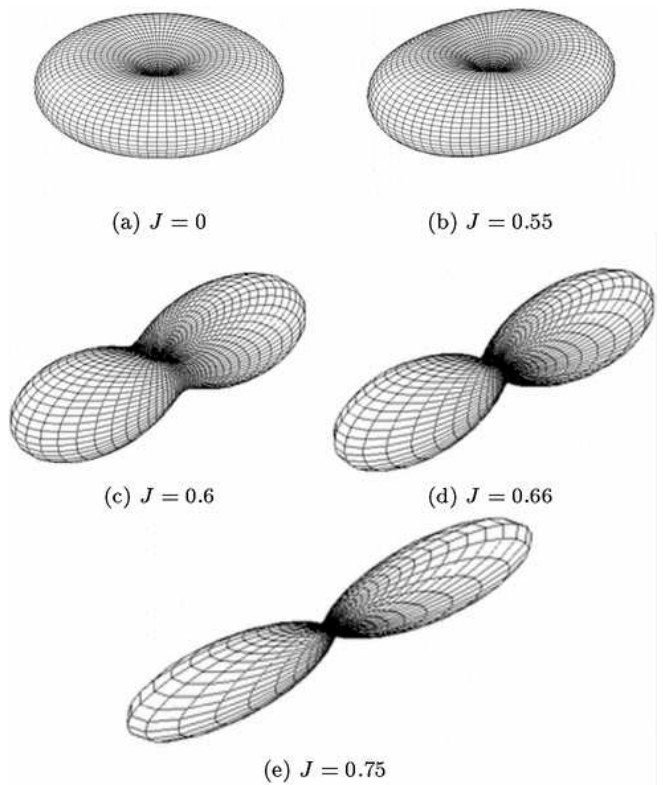}
\caption{\label{O3b2}Baby Skyrmions on the two-sphere ($\kappa^2=0.01$): 
The charge distribution $\mathcal{B}(\theta, \varphi)$ of the charge-two Skyrmion for different angular momenta.
In the figure, the vector 
$\mathcal{B}(\theta, \varphi) \vecr$
is plotted for the various $\theta$ and $\varphi$ values.$\hfill$}
\end{figure}

SBRS is also observed in rotating charge-two Skyrmions.
The static charge-two Skyrmion has only axial symmetry (Fig. \ref{O3b2}a),
with its symmetry axis having no preferred direction. 
Nonzero angular momentum aligns the axis of symmetry with the axis of rotation.
For small values of angular momentum, the Skyrmion is slightly deformed but remains axially symmetric (Fig. \ref{O3b2}b).
Above $J_{\textrm{crit}} \approx 0.55$ however, its rotational 
symmetry is broken, and it starts splitting to its `constituent' 
charge-one Skyrmions (Fig.  \ref{O3b2}c and \ref{O3b2}d). 
As the angular momentum is further increased, the splitting becomes more evident,
and the Skyrmion assumes a string-like shape.  This is somewhat reminiscent of the
well-known elongation, familiar from high-spin hadrons which are also known to assume 
a string-like shape with the constituent quarks taking position at the ends of the string \cite{Nambu, Kang}.

\begin{figure}[ht!]
\includegraphics[angle=0,scale=1,width=0.5\textwidth]{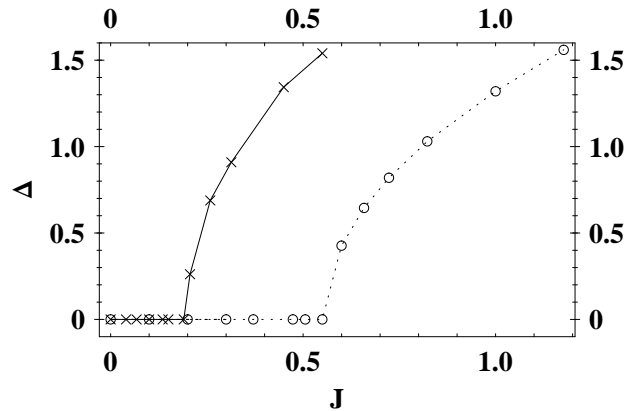}
\caption{\label{O3D2}The deviation from rotational symmetry $\Delta$ of the rotating charge-one (denoted by $\times$)
and charge-two (denoted by $\circ$) Skyrmions 
for different values of angular momentum. The lines are to guide the eye.$\hfill$}
\end{figure}

A quantitative measure for the deviation from rotational symmetry of the rotating solutions
may be obtained by evaluating the expression
\bea
\Delta^2= \int\left( \frac1{2 B} \int \mathcal{B}(\theta,\varphi) \sin \theta \rmd \theta \right)^2 \frac{\rmd \varphi}{2 \pi}   -1 \,,
\eea
where $\mathcal{B}(\theta,\varphi)$ is the charge density of the Skyrmion.
For rotationally-symmetric configurations $\Delta=0$. In Fig. \ref{O3D2},
$\Delta$ is plotted against the angular momentum $J$,
for both the charge-one and the charge-two solutions. 
The qualitative similarity to the bifurcation occurring 
in the rotating liquid mass system shown in Fig. \ref{gravEllips} is clear.

\subsection{The rational map ansatz}
A somewhat more analytical analysis of this system may be achieved by the use 
of the rational maps approximation scheme \cite{Rmaps3DSk1}, which is known to provide
quite accurate results for the static solutions of the model \cite{2Sphere3}.
In this approximation, points on the base sphere are expressed by the Riemann coordinate
$z=\tan \frac{\theta}{2} e^{i \varphi}$, and 
the ansatz for the field triplet is
\bea \label{eq:Rmap}
\bphi=( \frac{R+ \bar{R}}{1+|R|^2}, i \frac{R- \bar{R}}{1+|R|^2}, \frac{1-|R|^2}{1+|R|^2}) \,,
\eea
where the complex-valued  function $R(z)$ is a rational map of
degree $B$ between Riemann spheres
\bea
R(z)=\frac{p(z)}{q(z)} \,.
\eea
Here, $p(z)$ and $q(z)$ are polynomials in $z$, such that
$\max[\mbox{deg}(p),\mbox{deg}(q)]=B$,  and $p$ and $q$ have no common factors.  
Rational maps of degree $B$ correspond to field configurations with charge $B$. 
\par
In its implementation here, we have simplified matters even more and reduced the degrees of freedom
of the maps by a restriction only to 
those maps which exhibit the symmetries observed in the rotating full-field solutions.
This allowed the isolation of those parameters which are the most critical for
the minimization of the energy functional.
\par
\begin{figure}[ht!]
\subfloat[]{
\label{O3RmapB1} 
\includegraphics[angle=0,scale=1,width=0.5\textwidth]{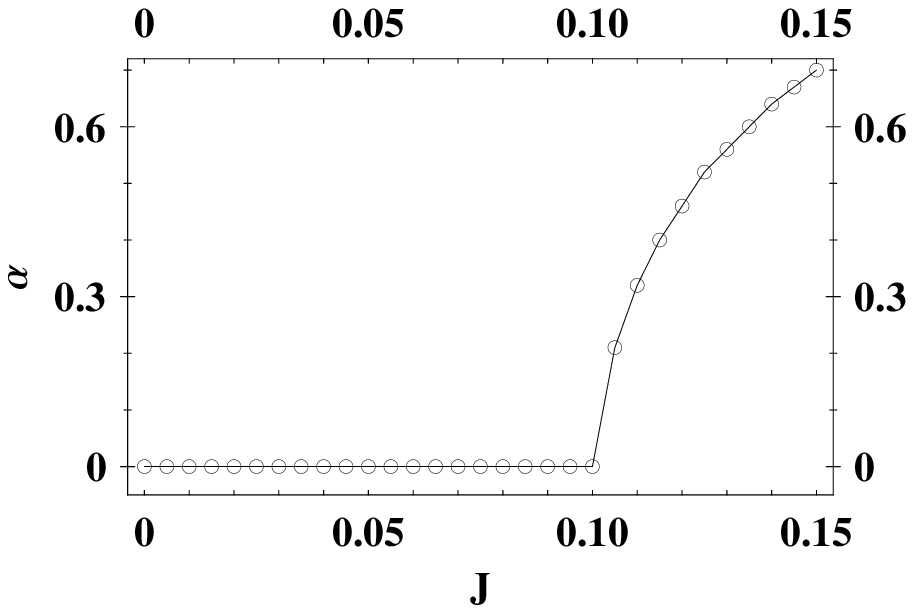}}
\hspace{1cm}
\subfloat[]{
\label{O3RmapB2} 
\includegraphics[angle=0,scale=1,width=0.5\textwidth]{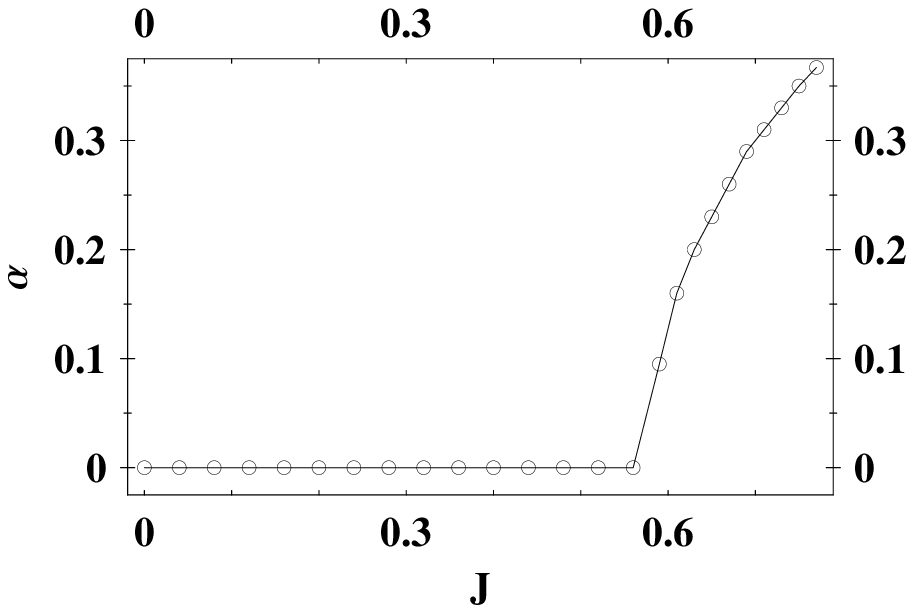}}
\caption{\label{O3Rmap}Spontaneous breaking of rotational symmetry 
in the restricted rational maps approximation for the baby Skyrmions on the two-sphere:
the parameter $\alpha$ as a function of the angular momentum $J$,
for the charge-one (top) and the charge-two (bottom) solutions.
The lines are to guide the eye.$\hfill$}
\end{figure}

In the charge-one rotating solution, the charge and energy densities exhibit two spatial symmetries. 
One is a reflection through the $xy$ plane 
(the plane perpendicular to the axis of rotation) and the other is a reflection through 
one horizontal axis. 
Enforcing these symmetries on rational maps of degree one 
results in the one-parametric family of rational maps
\bea \label{eq:RmapB1}
R(z)=\frac{\cos \alpha}{z+ \sin \alpha} \,, 
\eea
which produces the charge density
\bea
\mathcal{B}(\theta, \varphi)=\left( \frac{\cos \alpha}{1 +\sin \alpha \sin \theta \sin \varphi} \right)^2 \,, 
\eea
where $\alpha \in [-\pi, \pi]$ is the parameter of the map, and $\alpha=0$ corresponds to a 
rotationally-symmetric solution.
Results of a numerical minimization of the energy functional (\ref{eq:Eomega}) for fields constructed 
from (\ref{eq:RmapB1})
for different values of angular momentum $J$ are shown in
Fig. \ref{O3RmapB1}.
While for angular momentum less than $J_{\textrm{crit}} \approx 0.1$,
$\alpha=0$ minimizes the energy functional (a rotationally symmetric solution),
above this critical value bifurcation occurs and $\alpha=0$ is no longer a minimum;
the charge-one Skyrmion becomes ellipsoidal. 
\par
A similar analysis of the charge-two rotating solution yields the one-parametric map
\bea \label{eq:RmapB2}
R(z)=\frac{\sin \alpha +z^2 \cos \alpha}{\cos \alpha +z^2 \sin \alpha} \,,
\eea
with corresponding charge density
\bea
\mathcal{B}(\theta, \varphi)= \left( \frac{2 \cos 2 \alpha \sin \theta}
{2 + \sin^2 \theta  (\sin 2 \alpha \cos 2 \varphi-1)} \right)^2 \,. 
\eea
The results in this case are summarized in Fig. \ref{O3RmapB2},
indicating that above $J_{\textrm{crit}} \approx 0.57$ the minimal energy configuration
is no longer rotationally symmetric. 

\par
The discrepancies in the critical angular momenta $J_{\textrm{crit}}$
between the full-field method ($0.2$ for charge-one and $0.55$ for charge-two) 
and the rational maps scheme ($0.1$ for charge-one and $0.57$ for charge-two) 
are of course expected,
as in the latter method, the solutions have only one degree of freedom.
Nonetheless, the qualitative similarity in the behavior of the solutions 
in both cases is strong. 

\section{\label{sec:disk} The baby Skyrme model on a disk}
\begin{figure}[ht!]
\includegraphics[angle=0,scale=1,width=0.5\textwidth]{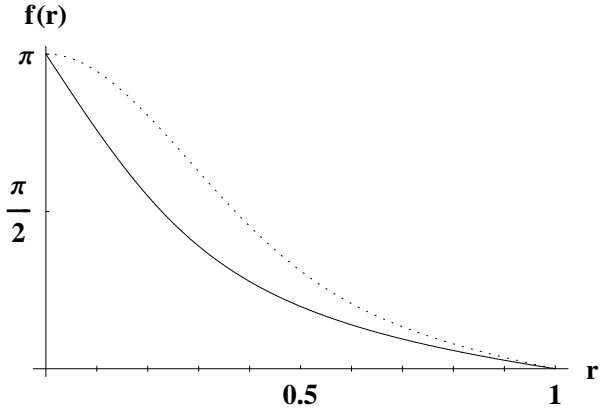}
\caption{\label{Prof12}The baby Skyrme model on a disk ($\mu^2=1$ and $\kappa^2=0.01$): 
Profile functions of the static charge-one (solid line) and charge-two (dotted line) Skyrmions.$\hfill$}
\end{figure}
\begin{figure}[ht!]
\includegraphics[angle=0,scale=1,width=0.48\textwidth]{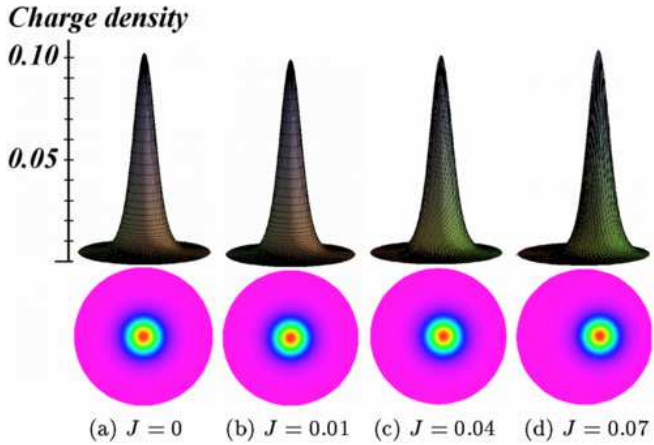}
\caption{\label{circleB1}Top: The charge density of the charge-one Skyrmion 
for different values of the angular momentum. Bottom: Corresponding contour plots  ranging from
violet (low density) to red (high density).
Above $J \approx 0.03$, the minimal energy configurations are no longer 
rotationally-symmetric; the center of mass of the Skyrmion is 
slightly shifted towards the bounding circle.$\hfill$}
\end{figure}
A second model in which SBRS is observed
is a baby Skyrme model for which radiation is inhibited by confining
the Skyrmion to the inside of a unit circle.
The domain $\mathbb{R}^2$ of the usual baby Skyrme model is  replaced by the unit disk 
\bea
D^2=\{\bx \in \mathbb{R}^2 : |\bx|^2 \leq 1 \} \,.
\eea
To recover the topology necessary for the existence of non-trivial solutions,
we require that the fields are the same in each direction on the bounding unit circle.
This results in the domain $D^2$ becoming topologically equivalent to a two-sphere,
and the topological charge is now given by the expression
\bea
\label{degree}
B = \frac1{ 4 \pi}\int_{D^2} \rmd r \rmd \varphi \, \bphi \cdot \left(\partial_r \bphi \times 
\partial_{\varphi} \bphi \right) \,,
\eea
where $r$ and $\varphi$ are the usual polar coordinates. 
As in the usual baby Skyrme model, the static solutions are found by minimizing the static energy
functional
\bea \label{eq:EstatD2}
E_{\textrm{static}}
&=&\frac1{4 \pi B}  \int_{D^2} r \rmd r \rmd \varphi \Big(
\frac1{2} (\partial_r \bphi \cdot \partial_r \bphi 
+ \frac1{r^2}\partial_{\varphi} \bphi \cdot \partial_{\varphi} \bphi) \nonumber\\
&+&\frac{\kappa^2}{2} \frac{(\partial_r \bphi \times \partial_{\varphi} \bphi)^2}{r^2}
+\mu^2 (1-\phi_3)
\Big) \,,
\eea
where the integration is over the unit disk, and the rotating solutions
are equivalently obtained by minimizing the functional
\begin{eqnarray} \label{Pot}
E_{\textrm{J}} &= &E_{\textrm{static}}+\frac{J^2}{2 I} \,,
\end{eqnarray}
where as before $I$ is the moment of inertia:
\bea \label{Moi}
I=\frac1{4 \pi B} \int_{D^2} r \rmd r \rmd \theta \left(
\partial_{\varphi} \bphi \cdot \partial_{\varphi} \bphi
+\kappa^2 (\partial_r \bphi \times \partial_{\varphi} \bphi)^2
\right) \,.
\eea
The numerical minimization of the energy functional has been 
carried out using the relaxation method discussed earlier in the case
of the baby Skyrme model on the two-sphere,
and the parameters of the model were fixed at $\mu^2=1$ and $\kappa^2=0.01$ for simplicity. 
\par
Here we focused our attention on the charge-one and charge-two Skyrmions,
as in the static limit these are found to be rotationally-symmetric with the form
\bea
\label{hedge}
\bphi(r,\theta) = \left(\sin f(r) \cos B\varphi,\sin f(r) \sin B\varphi , \cos f (r) \right),
\eea
where the profile function $f(r)$ satisfies the boundary conditions 
$f(0)= \pi$ and $f(1)=0$. Figure \ref{Prof12} shows the profile function 
obtained for each of the charges.

\begin{figure}
\includegraphics[angle=0,scale=1,width=0.48\textwidth]{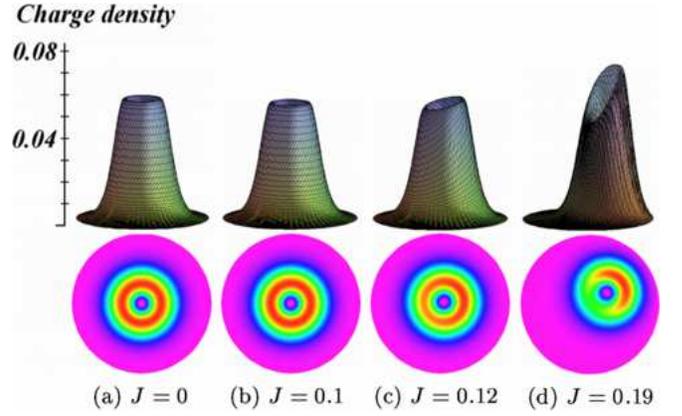}
\caption{\label{circleB2}Top: The charge density of the charge-two Skyrmion 
for different values of the angular momentum. Bottom: Corresponding contour plots  ranging from
violet (low density) to red (high density).
Above $J \approx 0.11$, the minimal energy configurations are no longer 
rotationally-symmetric.$\hfill$}
\end{figure}

As in the baby Skyrme model on the 
two-sphere, spontaneous breaking of rotational symmetry
is observed in this model as well.
In the charge-one sector, below $J \approx 0.03$ the stable solutions are rotationally-symmetric with only slight 
deformations from the static shape. Above this value SBRS appears;
the Skyrmion's center of mass shifts towards the bounding circle.
This is summarized in Fig. \ref{circleB1}.
\begin{figure}[ht!]
\includegraphics[angle=0,scale=1,width=0.5\textwidth]{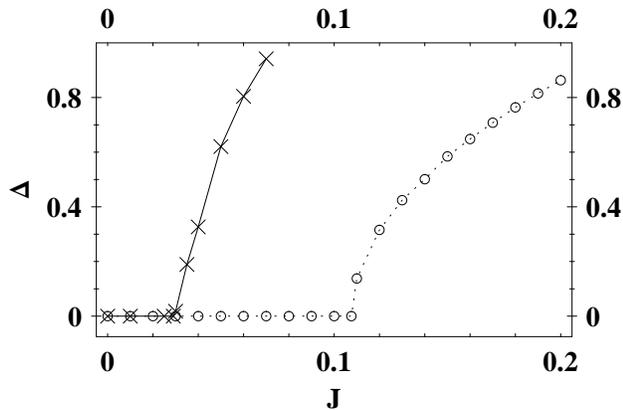}
\caption{\label{DiskD2}The baby Skyrme model on a disk: The symmetry-breaking measure $\Delta$ as a function of 
the angular momentum
for the charge-one (denoted by $\times$) and the charge-two (denoted by $\circ$) Skyrmions.$\hfill$}
\end{figure}
A similar situation occurs for the charge-two Skyrmions. 
The critical value there is  $J \approx 0.14$ as illustrated in Fig. \ref{circleB2}.
The behavior of the rotating solutions may be understood as follows;
by moving away from the center of the circle, the moment of inertia of the Skyrmion increases as dictated 
by Steiner's theorem. 
Since its shape remains more or less the same, 
its `self-energy' stays relatively unaffected (this is more evident in the charge-one case).

As with the baby Skyrmions on the two-sphere,
the deviation from rotational symmetry is measured by
\bea
\Delta^2= \int\left( \frac1{2 B} \int \mathcal{B}(r,\varphi) r \rmd r \right)^2 \frac{\rmd \varphi}{2 \pi}   -1 \,,
\eea
with $\mathcal{B}(r,\varphi)$ being the charge density of the Skyrmion.
In Fig. \ref{DiskD2}, $\Delta$ is plotted as a function of the angular momentum,
showing the emergence of SBRS as bifurcation points at the critical values of angular momentum.
\section{Summary and further remarks}
In this work we have studied spontaneous breaking of rotational symmetry (SBRS)
in two solitonic models whose solutions exhibit SBRS when the angular momentum is sufficiently high.
We have shown that the emergence of SBRS in these models 
can be directly linked to its appearance in classical mechanical systems,
such as the rotating liquid mass,
and that this linkage originates from general principles, and hence points out 
to the universality of this phenomenon.
\par
We believe that the results obtained in the present work may, at least
to some extent, also be linked to recent advances in the understanding
the non-sphericity of 
excited nucleons with of large orbital momentum.  
Non-spherical deformation of the
nucleon shape
is now a focus of considerable interest, both 
experimental \cite{exp1,exp2} and theoretical \cite{Miller1,Miller2,
Miller3}.
As Skyrmions are
known to provide a good qualitative description of many nucleon properties, 
we
hope that the results presented here will provide some corroboration to
recent results on this subject, {\it e.g.} ~\cite{Miller3}, although a more detailed
analysis of this analogy is in order.  We hope to be able to report on
these matters in forthcoming publications.
\hfill\break
\begin{acknowledgments} 
This work was supported in part by a grant from the Israel Science
Foundation administered by the Israel Academy of Sciences and Humanities.
\end{acknowledgments}


\begin{thebibliography}{100}
\bibitem{diffApp}
A different approach to the more general problem 
of a self-gravitating liquid mass of an arbitrary shape can be found in \cite{Lyttle}. 
\vfill\eject
\bibitem{pot}
The exact form of the potential in the baby Skyrme Lagrangian may be chosen almost arbitrarily, 
but we shall use the given potential as a reference case.
\end{thebibliography}
\end{document}